\documentclass{article}

\usepackage{microtype}
\usepackage{graphicx}
\usepackage{subfigure}
\usepackage{booktabs} 
\usepackage{adjustbox} 
\usepackage{multirow}  
\usepackage{threeparttable}

\usepackage{tikz}
\usetikzlibrary{calc}
\newcommand*\circled[1]{\tikz[baseline=(char.base)]{
    \node[shape=circle, draw, inner sep=0.02pt, yshift=0.07pt, 
        minimum height={1pt},] (char) {\vphantom{WAH1g}#1};}}

\usepackage{hyperref}

\usepackage[accepted]{mlsys2023}

\mlsystitlerunning{A Unified Compression Framework for Efficient Speech-Driven Talking-Face Generation}

\begin{document}

\twocolumn[
\mlsystitle{A Unified Compression Framework\\for Efficient Speech-Driven Talking-Face Generation}

\begin{mlsysauthorlist}
\mlsysauthor{Bo-Kyeong Kim}{nota}
\mlsysauthor{Jaemin Kang}{cnu}
\mlsysauthor{Daeun Seo}{cnu}
\mlsysauthor{Hancheol Park}{nota}
\mlsysauthor{Shinkook Choi}{nota}
\mlsysauthor{Hyoung-Kyu Song}{nota}
\mlsysauthor{Hyungshin Kim}{cnu}
\mlsysauthor{Sungsu Lim}{cnu}
\end{mlsysauthorlist}

\mlsysaffiliation{nota}{Nota Inc.}
\mlsysaffiliation{cnu}{Chungnam National University}

\mlsyscorrespondingauthor{Bo-Kyeong Kim}{bokyeong1015@gmail.com}
\mlsyscorrespondingauthor{Hyungshin Kim}{hyungshin@cnu.ac.kr}
\mlsyscorrespondingauthor{Sungsu Lim}{sungsu@cnu.ac.kr}

\mlsyskeywords{Talking-Face Generation, Efficient, Residual Block Removal, Knowledge Distillation, Mixed Precision Quantization, GAN Compression}

\vskip 0.3in

\begin{abstract}
Virtual humans have gained considerable attention in numerous industries, e.g., entertainment and e-commerce. As a core technology, synthesizing photorealistic face frames from target speech and facial identity has been actively studied with generative adversarial networks. Despite remarkable results of modern talking-face generation models, they often entail high computational burdens, which limit their efficient deployment. This study aims to develop a lightweight model for speech-driven talking-face synthesis. We build a compact generator by removing the residual blocks and reducing the channel width from Wav2Lip, a popular talking-face generator. We also present a knowledge distillation scheme to stably yet effectively train the small-capacity generator without adversarial learning. We reduce the number of parameters and MACs by 28× while retaining the performance of the original model. Moreover, to alleviate a severe performance drop when converting the whole generator to INT8 precision, we adopt a selective quantization method that uses FP16 for the quantization-sensitive layers and INT8 for the other layers. Using this mixed precision, we achieve up to a 19× speedup on edge GPUs without noticeably compromising the generation quality. 

\end{abstract}
]


\printAffiliationsAndNotice{}  

\section{Introduction}
\begin{figure}[t]
  \centering
  \includegraphics[width=\linewidth]{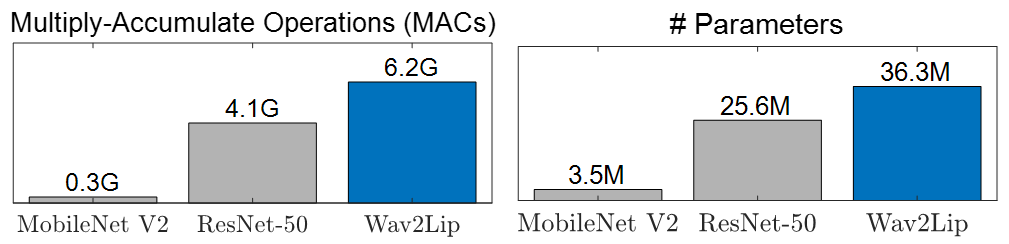}
  \caption{Computational comparison between classification and talking-face generation networks.} 
  \label{fig1_compCompute}
\end{figure}

Synthesizing face frames from target speech and facial identity has been actively studied with neural networks \cite{prajwal2020lip, wang2021facevid2vid, zhou2021pcavs, Song_2022_CVPR}. It enables a wide range of applications, e.g., digital human creation for entertainment industries and lip synchronization of dubbed videos. Despite impressive results of recent talking-face generation models, they are often computationally intensive, which can inhibit their practical deployment on resource-hungry devices. For instance, Wav2Lip \cite{prajwal2020lip} demands much heavier computations than well-known classification models (see \autoref{fig1_compCompute}).

Modern talking-face generation methods have been built upon generative adversarial networks (GANs) \cite{goodfellow2014generative}, which can provide visually plausible images. Recent studies toward efficient GANs have exploited knowledge distillation (KD) over pruned generators \cite{liu2021content, li2022learning, li2021revisiting}, neural architecture search \cite{li2020gan, fu2020autogan, Lin_2021_CVPR}, and quantization \cite{wang2019qgan, wan2020deep, andreev2021quantization}. These studies have focused merely on compressing classical image-to-image translation models (e.g., Pix2Pix \cite{pix2pix2017} and CycleGAN \cite{zhu2017unpaired}) rather than talking-face generators that often have more diverse architectural components and training objectives.

\begin{figure*}[h]
  \centering
  \includegraphics[width=\linewidth]{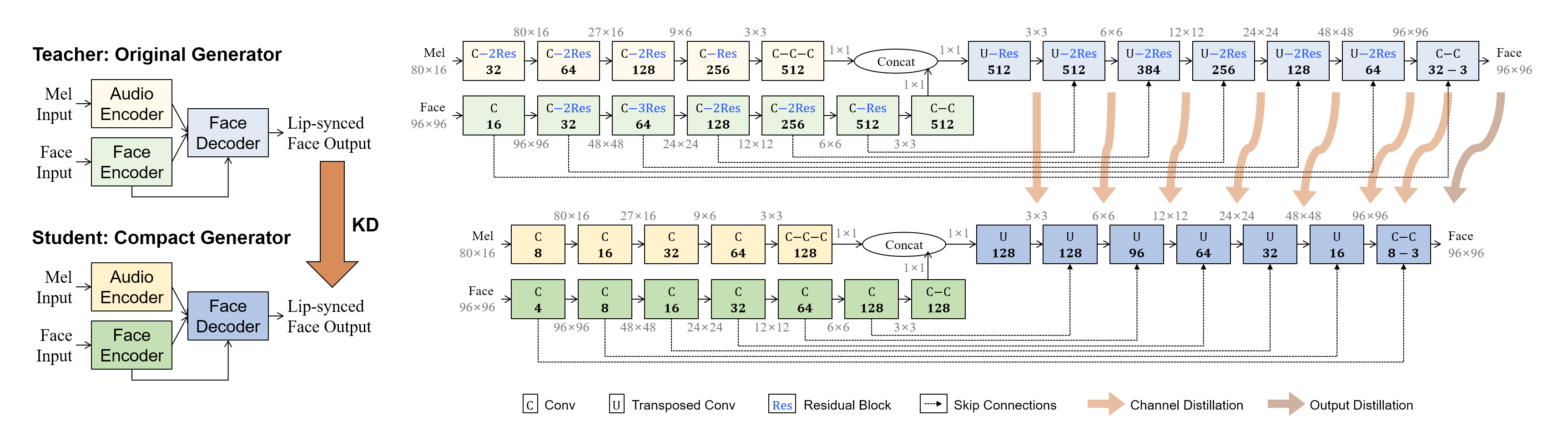}
  \caption{Generator architectures and KD process. Each layer is denoted by the type of convolution and the number of output channels. The compact generator with ×0.25 number of channels and removed residual blocks is trained under the guidance of the original generator.}
  \label{fig2_arch_kd}
\end{figure*}

This study presents a unified compression framework for efficient speech-driven talking-face synthesis by compressing Wav2Lip \cite{prajwal2020lip}\footnote{We chose Wav2Lip as our baseline because of its popularity and the existence of publicly available codes, allowing us to focus primarily on developing compression pipelines.}. First, we design a compressed architecture by reducing the number of channels and removing the residual blocks from the Wav2Lip generator. Then, we introduce an effective KD method that does not involve adversarial learning, thereby circumventing the challenge of preserving the Nash equilibrium between the discriminator and small-size generator. Experiments on the LRS3 dataset \cite{data_lrs3} show that our approach can compress the number of parameters and Multiply Accumulate Operations (MACs) by more than 28× while maintaining the generation quality of the original model. Moreover, we achieve 8×$\sim$17× inference speedups at FP32 and FP16 precision on NVIDIA Jetson edge GPUs. 

Furthermore, to overcome a severe performance drop when converting the whole generator to INT8 precision, we adopt a selective quantization method that utilizes FP16 for the quantization-sensitive layers and INT8 for the other layers. This mixed-precision method yields a 19× speedup on Jetson Xavier NX without losing the generation quality.

\section{Proposed Compression Framework}
We present an efficient talking-face generator obtained by compressing Wav2Lip \cite{prajwal2020lip} with three stages: designing a compact generator, effectively training it with KD, and employing a mixed-precision quantization.

\subsection{Compact Generator Architecture}
The Wav2Lip generator consists of two encoders, which process speech segments and face frames, and one decoder, which synthesizes lip-synced faces, as shown in \autoref{fig2_arch_kd}. We design a compressed architecture with two steps. First, we reduce the number of convolutional filters to become one-fourth of the original number. Second, we remove all the residual blocks from the original generator. We hypothesize that, because the model already receives a wealth of face-related information from the input, the residual blocks may be redundant and solely using the standard convolutions may be sufficient to fulfill the synthesis task.

\subsection{Knowledge Distillation (KD)}

A challenge in GAN compression is to identify an adequate discriminator capacity that can maintain the Nash equilibrium with the small-size generator \cite{li2021revisiting, ren2021omgd}. To sidestep this capacity imbalance issue, we employ a KD technique that does not require adversarial learning for training the small generator. 

The teacher, which is the original large generator, was pretrained with the following objective and became frozen:
\begin{equation} \label{loss_teacher}
\mathcal{L}^{Tea} = \lambda_{GAN}\mathcal{L}_{GAN}+\lambda_{Recon}\mathcal{L}_{Recon}+\lambda^{Tea}_{Sync}\mathcal{L}^{Tea}_{Sync},
\end{equation}

\noindent where the adversarial loss $\mathcal{L}_{GAN}$ to improve the visual fidelity, the reconstruction loss $\mathcal{L}_{Recon}$ to minimize the distance between the generated and ground-truth frames, and the lip-sync loss $\mathcal{L}^{Tea}_{Sync}$ to penalize inaccurate lip-sync results are identically defined as Eqs. (4), (2), and (3) of \citet{prajwal2020lip}. The weights $\lambda_{GAN}$, $\lambda_{Recon}$, and $\lambda^{Tea}_{Sync}$ are set as 0.07, 0.9, and 0.03, respectively.

The student, which is the compact generator, is trained using several distillation losses to have similar intermediate features and outputs as the teacher along with the lip-sync loss $\mathcal{L}^{Stu}_{Sync}$. The total student objective is computed as:

\begin{equation} \label{loss_student}
\mathcal{L}^{Stu} = \lambda_{CD}\mathcal{L}_{Ch-KD}+\mathcal{L}_{Out-KD}+\lambda^{Stu}_{Sync}\mathcal{L}^{Stu}_{Sync},
\end{equation}

\noindent where the channel KD loss $\mathcal{L}_{Ch-KD}$ to transfer intermediate feature information is equivalent to Eq. (10) of \citet{ren2021omgd}. The output KD loss $\mathcal{L}_{Out-KD}=\lambda_{SSIM}\mathcal{L}_{SSIM}+\lambda_{feature}\mathcal{L}_{feature}+\lambda_{style}\mathcal{L}_{style}+\lambda_{TV}\mathcal{L}_{TV}$, which is identical to Eq. (7) of \citet{ren2021omgd}, encourages the structural similarity using $\mathcal{L}_{SSIM}$ and the perceptual similarity using $\mathcal{L}_{feature}$ and $\mathcal{L}_{style}$ between the outputs of the student and the teacher and enforces the spatial smoothness in the student's outputs using $\mathcal{L}_{TV}$. The weights $\lambda_{CD}$, $\lambda_{SSIM}$, $\lambda_{feature}$, $\lambda_{style}$, $\lambda_{TV}$, and $\lambda^{Stu}_{Sync}$ are set as 10, 10, 10, 10000, 0.00001, and 3, respectively.

In the preliminary experiments, the offline KD (with the pretrained-and-frozen teacher) outperformed the online KD (with a simultaneous training of the teacher and the student) on our talking-face synthesis task, contrary to the results on image-to-image translation tasks reported in \citet{ren2021omgd}. Moreover, by comparing several locations to apply the channel KD loss, we found that KD over the last layers of the seven blocks in the face decoder performed well and KD over the encoder layers was deemed unnecessary.

\begin{table}[t]
\centering

\begin{adjustbox}{max width=0.835\columnwidth}
\begin{tabular}{c|ccc}
\hline
\multirow{2}{*}{Model} & \multicolumn{3}{c}{Precision} \\ \cline{2-4} 
                       & FP32 = FP16   & MIX   & INT8  \\ \hline
Teacher (original)     & 3.70           & 4.37  & 78.7  \\ \hline
Student (compressed)   & 5.30           & 5.06  & 157   \\ \hline
\end{tabular}
\end{adjustbox}

\caption{FID scores at different precisions. Lower is better. The INT8 quantization significantly degrades the generation quality, whereas the FP16-INT8 mixed-precision method (referred to as ``MIX") alleviates this issue.}

\label{table:fid_mix}
\end{table}

\begin{figure}[t]
  \centering
  \includegraphics[width=\linewidth]{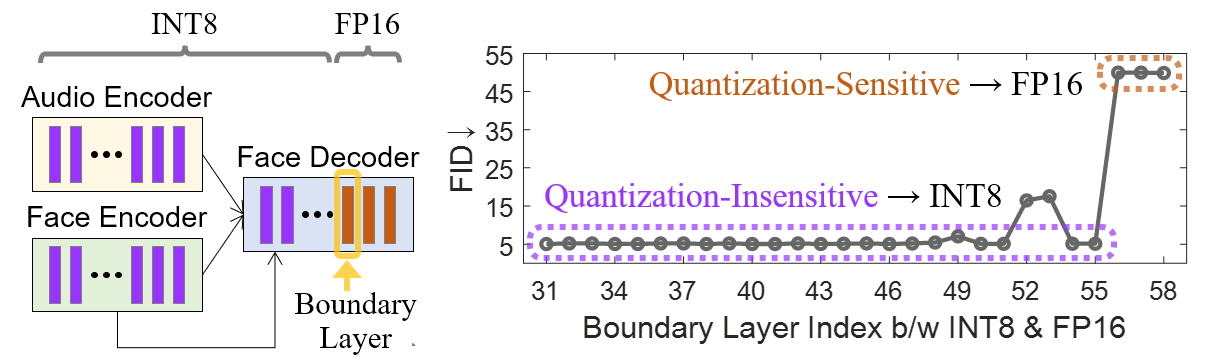}
  \caption{Layer-wise quantization sensitivity analysis for mixed-precision quantization.}
  \label{fig3_qnt_sensitivity}
\end{figure}

\begin{figure}[t]
  \centering
  \includegraphics[width=\linewidth]{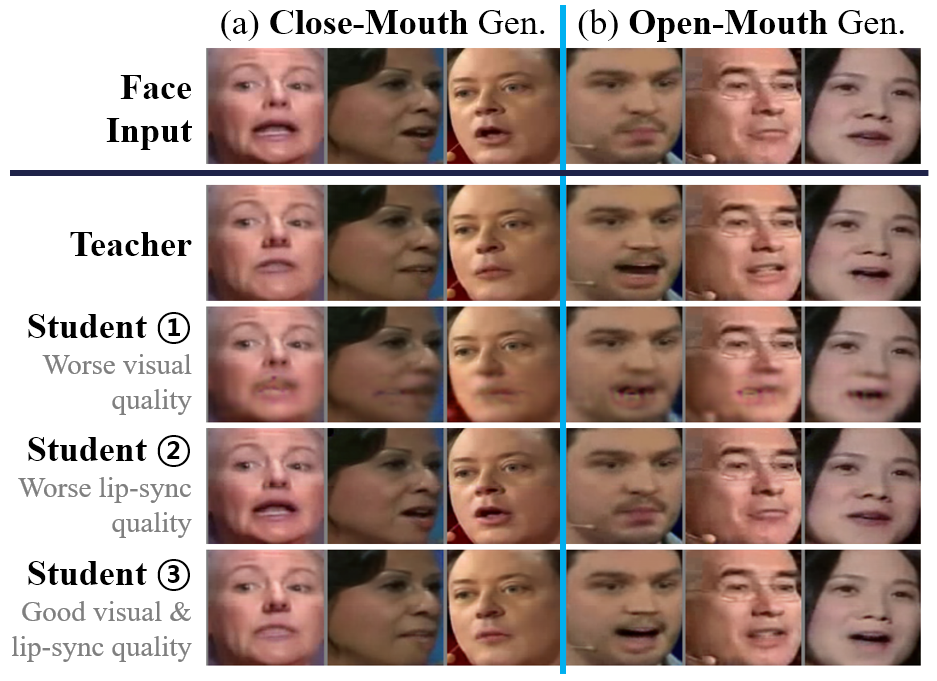}
  \caption{Qualitative results. As per the specified speech, the reference faces' mouth shapes should transform into (a) closed-lip and (b) open-lip shapes. The student models \circled{1}, \circled{2}, and \circled{3} correspond to those in \autoref{table:score}. The outputs of our final model (Student \circled{3}) closely resemble those of the original generator (Teacher).
  }
  \label{fig_visualResult}
\end{figure}

\subsection{Mixed-Precision Post-Training Quantization}
Quantization \cite{krishnamoorthi2018quantizing, nvidia_8bitQnt} enables the use of lower-precision representations in neural networks and improves computational efficiency. When converting floating-point operations to 8-bit integer (INT8) operations for the entire generator, we observed a significant degradation in the visual quality (see \autoref{table:fid_mix}). To overcome this issue, we adopt a hybrid-precision quantization approach \cite{cai2020zeroq, li2021brecq} that uses 16-bit floating-point (FP16) compute units for the quantization-sensitive layers and INT8 for the other layers. \autoref{fig3_qnt_sensitivity} shows a quantization sensitivity analysis that investigates the impact of switching the boundary layer index between INT8 and FP16 on FID performance through a layer-by-layer basis. We empirically find that applying FP16 precision to the decoder's output block performs well for the compact generator.\footnote{For the original large generator, applying FP16 to the first two encoder blocks and last two decoder blocks works well.}

\begin{figure*}[t]
  \centering
  \includegraphics[width=\linewidth]{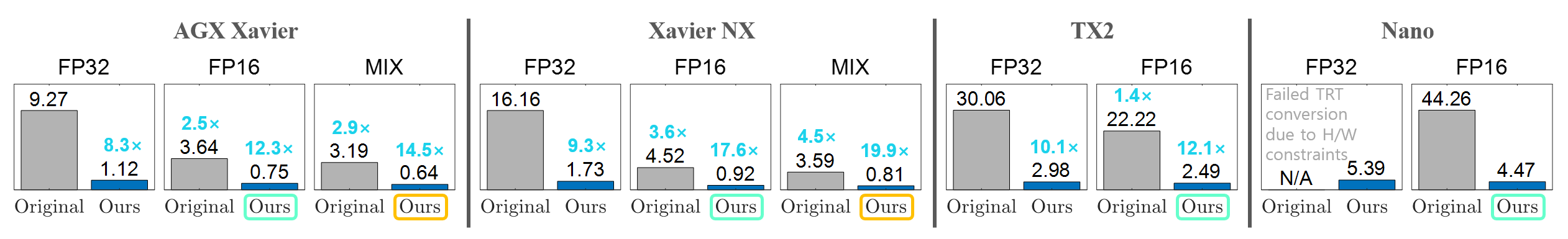}
  \caption{Latency (measured in milliseconds) at different precisions on NVIDIA Jetson edge GPUs. At FP16 precision, our approach boosts the inference speed by 8×$\sim$17×. At the mixed precision (denoted by ``MIX"), we achieve a 19× speedup on Xavier NX.}
  \label{fig_latency}
\end{figure*}

\section{Experimental Setup}

This section describes the datasets and evaluation metrics. See \autoref{appendix_impl} for the implementation details.

\subsection{Datasets}
Our approach is validated using the LRS3 dataset \cite{data_lrs3}, which consists of 32\textit{K} spoken sentences from TED clips, with the original \textit{train-val} and \textit{test} splits. As a calibration set for quantization, we use 10\textit{K} frames from the \textit{pre-train} split of the LRS2 dataset \cite{afouras2018lrs2}.

\subsection{Evaluation Metrics}
To evaluate the visual fidelity of generated frames, we use Fréchet Inception Distance (FID) \cite{heusel2017gans}. To evaluate the quality of lip synchronization between speech samples and generated face images, we compute Lip Sync Error - Distance (LSE-D) and Confidence (LSE-C) \cite{chung2016out, prajwal2020lip}. To evaluate the computation, we measure the actual latency on edge GPUs as well as the number of parameters and MACs.

\begin{table}[t]
\centering

\begin{adjustbox}{max width=1\columnwidth}
\begin{threeparttable}

\begin{tabular}{ccccc|ccc|cc}
\hline
\multicolumn{5}{c|}{Model}                                                                  & \multicolumn{3}{c|}{Performance\tnote{2}} & \multicolumn{2}{c}{Computation}                                \\ \hline
\begin{tabular}[c]{@{}c@{}}Type\\ (\# Ch.)\end{tabular}                   & \multicolumn{2}{c}{\begin{tabular}[c]{@{}c@{}}Removed\\ ResBlocks?\end{tabular}}   & \begin{tabular}[c]{@{}c@{}}Use\\ KD?\end{tabular} & \begin{tabular}[c]{@{}c@{}}Sync\\ Step\tnote{1} \end{tabular} & FID↓     & LSE-D↓    & LSE-C↑    & MACs                           & \# Params                     \\ \hline
\begin{tabular}[c]{@{}c@{}}Teacher\\ (×1.0)\end{tabular}                      & \multicolumn{2}{c}{X}                  & X  & Mid        & 3.70      & 6.48      & 7.78      & 6.21G                          & 36.3M                         \\ \hline
\multirow{6}{*}{\begin{tabular}[c]{@{}c@{}}Student\\ (×0.25)\end{tabular}} & \multicolumn{2}{c}{\multirow{3}{*}{X}} & X  & All        & 94.6     & 11.4      & 2.08      & \multirow{3}{*}{\begin{tabular}[c]{@{}c@{}}0.40G\\ (15.6×)\end{tabular}} & \multirow{3}{*}{\begin{tabular}[c]{@{}c@{}}2.3M\\ (15.9×)\end{tabular}} \\ \cline{4-8}
                                 & \multicolumn{2}{c}{}                   & X  & Mid        & 5.19     & 7.06      & 6.89      &                                &                               \\ \cline{4-8}
                                 & \multicolumn{2}{c}{}                   & O  & Mid        & 5.49     & 6.10       & 8.41      &                                &                               \\ \cline{2-10} 
                                 & \multirow{3}{*}{O}        & \circled{1}        & X  & All        & 23.1     & 7.34      & 6.28      & \multirow{3}{*}{\begin{tabular}[c]{@{}c@{}}0.22G\\ (\textbf{28.8×})\end{tabular}} & \multirow{3}{*}{\begin{tabular}[c]{@{}c@{}}1.3M\\ (\textbf{28.9×})\end{tabular}} \\ \cline{3-8}
                                 &                             & \circled{2}       & X  & Mid        & 4.17     & 11.4      & 2.67      &                                &                               \\ \cline{3-8}
                                 &                             & \circled{3}       & O  & Mid        & \textbf{5.30}      & \textbf{6.35}      & \textbf{8.04}      &                                &                               \\ \hline
\end{tabular}
\begin{tablenotes}
\item[1] The lip-sync loss is used during all training steps (All) or from the middle step (Mid).
\item[2] The symbols ↓ and ↑ denote that lower and higher values are preferable, respectively.
\end{tablenotes}

\end{threeparttable}
\end{adjustbox}
\caption{Quantitative evaluation on the LRS3 dataset. Our compressed generator, with ×0.25 channel numbers and removed residual blocks, reduces computation by over 28 times. The use of knowledge distillation stabilizes the training of the small generator, effectively addressing the tradeoff between visual fidelity (FID) and lip-sync quality (LSE-D and LSE-C). }

\label{table:score}
\end{table}

\begin{table}[]
\centering

\begin{adjustbox}{max width=0.98\columnwidth}
\begin{tabular}{c|ccc|cc}
\hline
\multirow{2}{*}{Method}                                       & \multicolumn{3}{c|}{Performance} & \multicolumn{2}{c}{Computation}                                                                                  \\ \cline{2-6} 
                                                              & FID↓     & LSE-D↓    & LSE-C↑    & MACs                                                    & \# Params                                              \\ \hline
\begin{tabular}[c]{@{}c@{}}Cut Inner Layers\\ \cite{kim2022cut}\end{tabular}          & 6.09     & 7.29      & 6.61      & \begin{tabular}[c]{@{}c@{}}0.70G\\ (8.9×)\end{tabular}  & \begin{tabular}[c]{@{}c@{}}1.9M\\ (18.9×)\end{tabular} \\ \hline
\begin{tabular}[c]{@{}c@{}}Lite Wav2Lip\\ (Ours)\end{tabular} & \textbf{5.30}     & \textbf{6.35}      & \textbf{8.04}      & \begin{tabular}[c]{@{}c@{}}0.22G\\ (\textbf{28.8×})\end{tabular} & \begin{tabular}[c]{@{}c@{}}1.3M\\ (\textbf{28.9×})\end{tabular} \\ \hline
\end{tabular}
\end{adjustbox}
\caption{Comparison to the previous method \cite{kim2022cut} for compressing Wav2Lip on the LRS3 dataset.}

\label{table:comp}
\end{table}

\section{Experimental Results}

\subsection{Quantitative Results}

\autoref{table:score} summarizes the quantitative results on the LRS3 dataset. The reduction of channel numbers for the small generator leads to a 15× reduction in computation, while the removal of residual blocks makes it even more efficient and brings a 28× reduction. The training of the small generator without KD causes a compromise between the lip-sync error and the visual quality.\footnote{In the absence of KD, we explore different ways of incorporating the lip-sync loss (marked as ``Sync Step" in \autoref{table:score}) to check if it would improve the results. Despite our attempts, the trade-off between both metrics remains unresolved: using the lip-sync loss for the entire training process (marked as ``All") yields moderate lip-sync errors but significantly sacrifices the visual fidelity; using it from the middle of training (marked as ``Mid") improves the generation quality but negatively impacts the lip-sync quality.} In contrast, the use of KD stabilizes the training process and mitigates this trade-off, achieving high performance in both metrics. 

\autoref{table:comp} shows the comparison with the previous method, Cut Inner Layers (CIL) \cite{kim2022cut}, based on structured pruning of inner layers. Our approach outperforms CIL in terms of both performance and computation.

\subsection{Visual Results}

\autoref{fig_visualResult} depicts some generated results. In accordance with the given target speech, the left faces' mouths should close and the right faces' mouths should open. Without KD, either the visual appearance or the lip-sync quality is unsatisfactory. However, KD enables the compact generator to produce accurately lip-synced face frames that match the quality of those from the original generator.

\subsection{Inference Speed on Edge GPUs}
We further demonstrate actual speed gains of our approach on edge GPUs belonging to the NVIDIA Jetson family: AGX Xavier, Xavier NX, TX2, and Nano. With TensorRT acceleration at FP32 and FP16 precision, our generator achieves 8.3×$\sim$17.6× inference speedups in comparison to the original model. With the mixed-precision quantization\footnote{To the best of our knowledge, INT8 operations were not supported in TX2 and Nano devices during the time of our research, and thus the mixed-precision results for these devices are not included in \autoref{fig_latency}.} that selectively uses INT8 or FP16 for individual layers, our generator exhibits a 19.9× speed improvement on Xavier NX and a 14.5× speedup on AGX Xavier without a noticeable decline in the generation quality. We remark that these results are better than the speedups obtained solely using FP16 precision for all the layers. \autoref{appendix_latency} presents additional results including the latency at INT8 precision.

\section{Conclusion}
This work introduces a unified framework toward efficient speech-driven talking-face generation and its application to Wav2Lip compression. Our compact generator with removed residual blocks is trained under well-designed knowledge distillation and is further optimized using mixed-precision quantization. We obtain 28× computational reduction while preserving the generation quality. We also show actual speedups on edge GPUs. Future research can explore an automatic way to determine the quantization precision of individual layers for compressing talking-face generators.

\section*{Acknowledgements}
We thank the NVIDIA Applied Research Accelerator Program for supporting this study.

\bibliographystyle{mlsys2023}
\bibliography{paper}


\newpage
\appendix

\section*{Appendix}

\section{Implementation Details} \label{appendix_impl}
We adopt the codes of Wav2Lip\footnote{https://github.com/Rudrabha/Wav2Lip} for constructing generator models and OMGD\footnote{https://github.com/bytedance/OMGD} for training them with KD. A single NVIDIA GeForce RTX 3090 GPU is utilized for training. We implement the selective quantization with torch2trt\footnote{https://github.com/NVIDIA-AI-IOT/torch2trt} and further optimize the inference on edge GPUs with NVIDIA TensorRT\footnote{https://developer.nvidia.com/tensorrt}.

\section{Latency on Edge GPUs} \label{appendix_latency}
\autoref{table:appendix_comp} shows additional latency results at INT8 precision along with the results presented in \autoref{fig_latency} of the main text. In our experiments, the original large generator at FP32 precision is not deployable on Jetson Nano using TensorRT due to the hardware constraints. In contrast, our compressed generator can be deployed with the latency of only 5.39 ms. We remark that the INT8 uniform quantization yields slightly better latency results than the FP16-INT8 mixed-precision quantization (denoted by ``MIX") but considerably degrades the generation quality (see \autoref{table:fid_mix} of the main text).

Because INT8 operations were not supported in Jetson TX2 and Nano during the time of this study, the mixed-precision results for these devices are not included in \autoref{table:appendix_comp} and \autoref{fig_latency}.

\begin{table}[h]
\centering

\begin{adjustbox}{max width=0.95\columnwidth}

\begin{tabular}{cc|cccc}
\hline
\multicolumn{2}{c|}{Model}    & \multicolumn{4}{c}{NVIDIA Jetson Device}                                                                                                                                                                                                               \\ \hline
Prec.                 & Type  & AGX                                                      & NX                                                       & TX2                                                      & Nano                                                    \\ \hline
\multirow{2}{*}{FP32} & Original & 9.27ms                                                   & 16.16ms                                                  & 30.06ms                                                  & N/A                                                     \\ \cline{2-6} 
                      & Ours  & \begin{tabular}[c]{@{}c@{}}1.12ms\\ (8.3×)\end{tabular}  & \begin{tabular}[c]{@{}c@{}}1.73ms\\ (9.3×)\end{tabular}  & \begin{tabular}[c]{@{}c@{}}2.98ms\\ (10.1×)\end{tabular} & \begin{tabular}[c]{@{}c@{}}5.39ms\\ ($\infty$×)\end{tabular}  \\ \hline
\multirow{2}{*}{FP16} & Original & \begin{tabular}[c]{@{}c@{}}3.64ms\\ (2.5×)\end{tabular}  & \begin{tabular}[c]{@{}c@{}}4.52ms\\ (3.6×)\end{tabular}  & \begin{tabular}[c]{@{}c@{}}22.22ms\\ (1.4×)\end{tabular} & \begin{tabular}[c]{@{}c@{}}44.26ms\\ ($\infty$×)\end{tabular} \\ \cline{2-6} 
                      & Ours  & \begin{tabular}[c]{@{}c@{}}0.75ms\\ (\textbf{12.3×})\end{tabular} & \begin{tabular}[c]{@{}c@{}}0.92ms\\ (\textbf{17.6×})\end{tabular} & \begin{tabular}[c]{@{}c@{}}2.49ms\\ (\textbf{12.1×})\end{tabular} & \begin{tabular}[c]{@{}c@{}}4.47ms\\ (\textbf{$\infty$×})\end{tabular}  \\ \hline
\multirow{2}{*}{MIX}  & Original & \begin{tabular}[c]{@{}c@{}}3.19ms\\ (2.9×)\end{tabular}  & \begin{tabular}[c]{@{}c@{}}3.59ms\\ (4.5×)\end{tabular}  & \multirow{2}{*}{-}                                       & \multirow{2}{*}{-}                                      \\ \cline{2-4}
                      & Ours  & \begin{tabular}[c]{@{}c@{}}0.64ms\\ (\textbf{14.5×})\end{tabular} & \begin{tabular}[c]{@{}c@{}}0.81ms\\ (\textbf{19.9×})\end{tabular} &                                                          &                                                         \\ \hline
\multirow{2}{*}{INT8} & Original & \begin{tabular}[c]{@{}c@{}}2.38ms\\ (3.9×)\end{tabular}  & \begin{tabular}[c]{@{}c@{}}3.13ms\\ (5.2×)\end{tabular}  & \multirow{2}{*}{-}                                       & \multirow{2}{*}{-}                                      \\ \cline{2-4}
                      & Ours  & \begin{tabular}[c]{@{}c@{}}0.63ms\\ (14.8×)\end{tabular} & \begin{tabular}[c]{@{}c@{}}0.74ms\\ (21.8×)\end{tabular} &                                                          &                                                         \\ \hline
\end{tabular}

\end{adjustbox}
\caption{Inference speed (measured in milliseconds) at various precisions on edge GPUs.}

\label{table:appendix_comp}
\end{table}

\end{document}